\documentclass[%
 aip,
 jmp,%
 amsmath,amssymb,
preprint,%
]{revtex4-1}

\usepackage{graphicx}
\usepackage{dcolumn}
\usepackage{bm}

\begin{document}


\title[]{Gauss law at a vertex in Lattice QCD and its gauge invariant Hilbert space.}
\author{T P Sreeraj}%
\email{sreeraj.tp@gmail.com}
\author{Ramesh Anishetty}
\email{ramesha@imsc.res.in}
\affiliation{ 
The Institute of Mathematical Sciences, C. I. T. campus, Taramani, Chennai
}%


\date{\today}

\begin{abstract}
	We solve the local Gauss law in lattice QCD in the presence of matter charges. This corresponds to constructing  singlet states using Schwinger Bosons and Fermions of SU(3) group at each vertex of the lattice. 
 We also calculate the action of various invariant operators on these states  required for studying the dynamics. 

\end{abstract}

\maketitle


\section{Introduction}
In lattice gauge theory, in the absence of matter fields, construction of gauge invariant Hilbert space can be reduced\cite{rs1}, by a process called point splitting, to the problem of constructing a 
 singlet Hilbert space at the vertex  which satisfies the Gauss law $(E_1^a+E_2^a+E_3^a)= 0$ where $E_1^a, E_2^a,E_3^a$ are the generators of SU(3) in the adjoint representation.  
One can construct \cite{rsarx} such a Hilbert space by using Schwinger Boson representation of SU(3) algebra. 
	 However, in the presence of matter fields, one has to further construct a singlet space which satisfies $(E_1^a+E_2^a+{\cal E}^a)= 0$ where $E_1^a,E_2^a$ are the chromoelectric fields in the adjoint representation and ${\cal E}^a$ is the color currents due to matter fields which can be in any representation. In particular in QCD, it is in the $3$ and $3^*$ fermionic representation. 
	 We construct such a singlet Hilbert space here. We also calculate the action of various invariant operators on the basis of the singlet Hilbert space so constructed. These actions are essential for studying the Hamiltonian of lattice QCD.

The plan of the paper is as follows. In section \ref{s:gl}, we describe Gauss law in lattice QCD in the presence of matter fields and motivate the problem. In section\ref{su2}, we address the problem in the context of simpler SU(2). Following a short description of Schwinger boson and fermionic representation of SU(2) algebra, we construct a singlet Hilbert space for SU(2). We then write down the action of various invariant operators on this basis. In section \ref{su3}, we repeat the same for SU(3). 
\section{Gauss law in lattice gauge theory}
\label{s:gl}
The dynamical variables of the Hamiltonian formulation \cite{lgt,prep} of SU(3) lattice gauge theory are the link operators $U_i(n)$, their conjugate electric fields $E_i(n)$  and fermionic(spinor) matter fields $q(n)$ and their conjugate variables $iq^\dagger(n)$. $U_i(n)$ is a $3\times 3$ operator valued matrix belonging to the SU(3) group lying on the link starting at site $n$ in the $i$ direction  and $E_i(n)$ are SU(3) lie algebra valued operators. The matter fields transform under the fundamental $3$ representation of SU(3) and its conjugate variable transforms under the $3^*$ representation. One can define a right electric field $E_{\bar i}(n+i)$ by parallel transporting $E_i(n)$ along the link $(n,i)$. $E_{\bar i}^a(n+i)=R^{a}_{~b}(U_i(n)) E_{i}^b(n); R^a_{~b}(U)= \frac{1}{2}Tr(U \lambda^a U^\dagger \lambda_b)$, where $\lambda^a$ are the Gellmann matrices. On a square lattice, the Gauss law at a site $n$ can be stated as follows: 
\begin{align}
{\cal G}^a(n)\equiv \sum\limits_{i=1,2} E_i^a(n)+E_{\bar i}^a(n)+ {\cal E}^a(n) =0
\end{align}
Here, we consider a $2+1$ dimensional lattice and ${\cal E}^a$ is the color current of matter field and is given by 
${\cal E}^a=q^\dagger\frac{\lambda^a}{2}q$ for fermion fields. ${\cal E}^a$ obeys SU(3) algebra. ${\cal G}^a(n)$ generates the gauge transformations at site $n$. 
\begin{figure}
	\centering
	\includegraphics[scale=1.5]{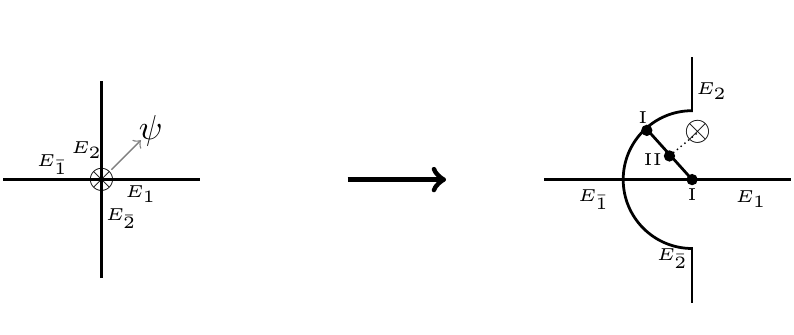}
	\caption{Point splitting for a $2+1$ dimensional gauge theory. A 4-vertex becomes two 3-vertices of type I and one 3 vertex of type II. $\otimes$ denotes a fermion.}
	\label{f:ps} 
\end{figure}

In $2+1$ dimensions, point splitting \cite{rs1} in the presence of matter fields replaces the gauss law at a 4-vertex on a square lattice by Gauss laws at three 3-vertices. It essentially involves introducing new links so that the lattice reduces to a collection of 3-vertices connected together by links. New link variables and electric fields are introduced on the new links and Gauss law constraints are imposed on all the 3-vertices. This has been shown \cite{rs1} to be equivalent to the original square lattice. The point splitting in $2+1$ dimensional lattice gauge theory with matter fields is graphically depicted in figure (\ref{f:ps}). In general, after point splitting, there are two types of 3-vertices which we call type I and II. Gauss law at the type I vertex is given by $(E_1^a+E_2^a+E_3^a)= 0$. This is the only type of vertex when there are no matter fields. As mentioned in the introduction, singlet space satisfying the above Gauss law at type I vertex has already been constructed \cite{rsarx}. The Gauss law at type II vertex is 
$(E_1^a+E_2^a+{\cal E}^a)= 0$. The construction of a Hilbert space which satisfies this constraint and the action of local, invariant operators on this Hilbert space will occupy the rest of the paper.
Note that here we are interested only in the construction of a singlet space and therefore have ignored the spinor indices on the fermion fields. Including spin degrees of freedom adds some more type II vertices which can be dealt with in similar manner as here. For simplicity, we will first discuss SU(2) fermions and then SU(3).
\section{SU(2)}
\label{su2}
Bosonic representation of SU(2) generators in terms of Schwinger bosons \cite{schwinger,prep} is given by: 
\begin{align}
E^{a}(\vec{x}) &\equiv 
a^{\dagger}_\alpha\left(\frac{\sigma^{a}}{2}\right)_{\alpha \beta}a_{\beta}
\end{align}
where $a^\dagger_\alpha, a_\beta$ are the creation, annihilation operator doublets of harmonic oscillators satisfying $[a_\alpha,a^\dagger_\beta]=\delta_{\alpha\beta}$.
The basis of any irreducible representation space of SU(2) can be created as 
\begin{align}
|j,m\rangle\equiv |n_1,n_2\rangle= \frac{(a^\dagger_1)^{n_1}(a^\dagger_2)^{n_2}}{n_1!n_2!}|0\rangle
\label{su2basis}
\end{align} 
Above, $j=\frac{n_1+n_2}{2},m=\frac{n_1-n_2}{2}$ are the eigen values of $E^2,E^{(3)}$ and $|0\rangle$ is the harmonic oscillator vacuum $a|0\rangle=0$.
$a^\dagger_\alpha$ transform under the fundamental representation of SU(2). 

A fermionic representation of the SU(2) generators can be constructed 
as follows 
\begin{align}
	{\mathcal E}^a = q^\dagger_\alpha \Big(\frac{\sigma^a}{2}\Big)_{\alpha\beta}q_\beta
\end{align}
Where $q,q^\dagger$ are fermionic creation annihilation doublet operators satisfying $\{q_\alpha,q^\dagger_\beta\}=\delta_{\alpha\beta}; \{q_\alpha,q_{\beta}\}=0; \{q^\dagger_\alpha,q^\dagger_{\beta}\}=0 $.
Fermionic creation operators create only the fundamental representation of SU(2).   \begin{align}
	|j=\frac{1}{2},m=\frac{1}{2}\rangle=|n_{q1}=1,n_{q2}=0\rangle = q^\dagger_1 |0\rangle \hspace{1cm}|j=\frac{1}{2},m=-\frac{1}{2}\rangle= |n_{q1}=0,n_{q2}=1\rangle=q^\dagger_2|0\rangle
\end{align}
Above, $n_{q1},n_{q2}$ are the eigenvalues of the number operators $N_{q1}=q^\dagger_1\cdot q_1,N_{q2}=q^\dagger_2\cdot q_2 $. These number operators are related to ${\cal E}^2,{\cal E}^{(3)}$ as ${\cal E}^2=3\Big(\frac{N_q}{2}\Big)\Big(1-\frac{N_q}{2}\Big),{\cal E}^{(3)}=N_{q1}-N_{q2}$ where $N_q=N_{q1}+N_{q2}.$
\begin{figure}
	\includegraphics{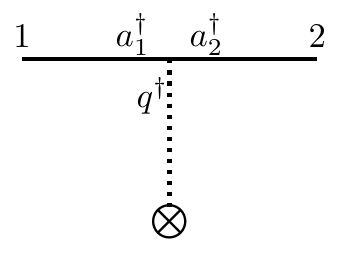}
	\caption{Type-II 3-vertex for SU(2). $a^\dagger_i$ denotes the prepotential doublets and $q^\dagger$ denotes the fermionic doublets at the vertex.}
	\label{3vertex}
\end{figure}

 Consider the direct product space of two arbitrary SU(2) irrep in the bosonic representation and a fundamental irrep in the fermionic representation  with a basis $|j_1,m_1\rangle|j_2,m_2\rangle$ and $|j_3=\frac{1}{2},m_3=\pm\frac{1}{2}\rangle$ respectively. We are interested in finding a subspace which satisfies the constraint $E_1^a+E_2^a+{\cal E}^a=0$ i.e, a subspace which transform as a singlet under SU(2). Such a representation is denoted pictorially in Figure. \ref{3vertex}. In order to construct a basis of such a singlet space we consider the action of all the invariant operators involving only creation operators on the oscillator vacuum $a_i|0\rangle=q|0\rangle=0$:
 \begin{align}
 	|s_{12},r_i,v \rangle_u= (a_1^\dagger \cdot \tilde{a}_2^\dagger )^{s_{12}} (a^\dagger_i \cdot \tilde{q}^\dagger)^{r_i}(q^\dagger \cdot \tilde{q}^\dagger)^v |0\rangle
 \end{align}
where, $s_{12}$ is a non-negative integer, $r_i,v =0,1$. Subscript $u$ is used for unnormalized states. 
However, all these states are not independent due to the following relations: 
	 \begin{align}(i)~~~ (a^\dagger_1 \cdot \tilde{q}^\dagger)(a^\dagger_2\cdot \tilde{q}^\dagger)&=\frac{1}{2} (a^\dagger_1\cdot \tilde{a}^\dagger_2)(q^\dagger \cdot \tilde{q}^\dagger) \nonumber\\
	 	(ii)~~~ (a^\dagger_i\cdot \tilde{q}^\dagger)(q^\dagger \cdot \tilde{q}^\dagger)&=0
		\end{align}
	These relations implies that certain states can be parametrized in multiple ways using the quantum numbers $s_{12},r_i,v$. For instance, $|s_{12},r_1=1,r_2=1,v=0\rangle_u = |r_1=0,r_2=0,v=1,s_{12}+1\rangle_u$. 
	A unique and complete labelling of states is achieved if we respect the condition: 
	\begin{align}
	 r_1r_2+r_1v+r_2v=0
	 \label{su2cond}
	 \end{align}
	  I.e, two quantum numbers among $r_1,r_2,v$ cannot be non-zero in a state. 
	

\begin{figure}
	\centering
	\includegraphics[scale=1.5]{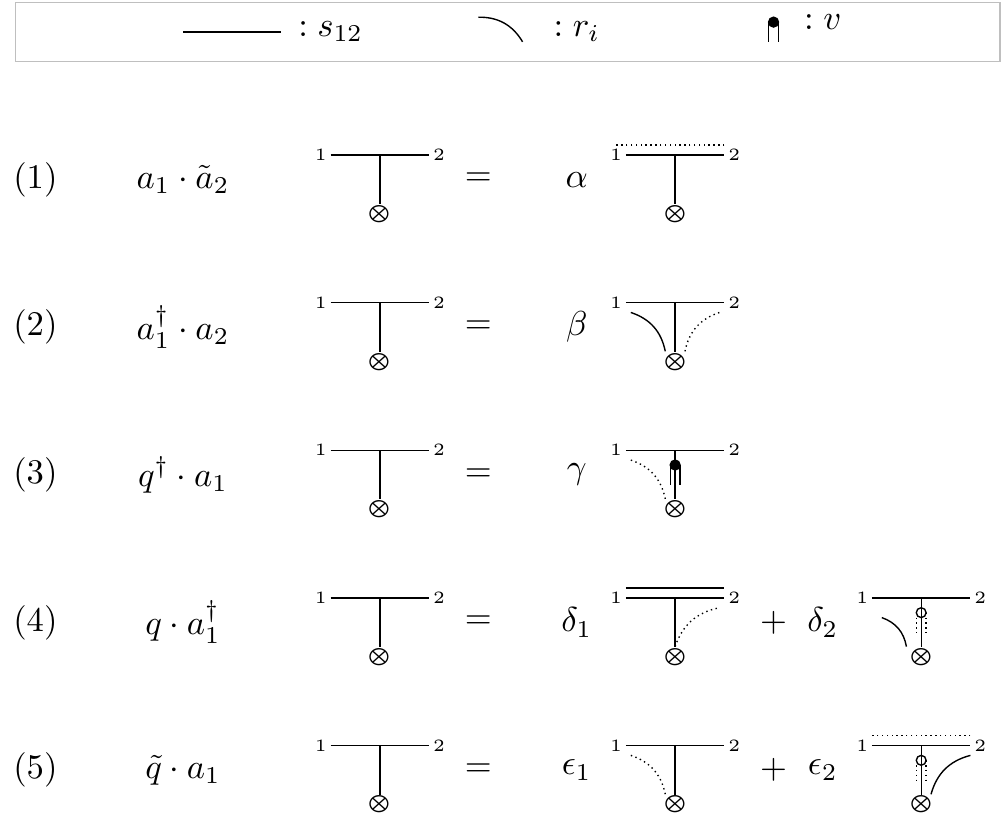}
	\caption{graphical representation of action of SU(2) invariant operators on the basis states. solid lines denote an increase in the corresponding quantum number by 1 and dotted lines denotes decrease by one.} 
\end{figure}
Action of various invariant operators on these unnormalized basis states as well as the norm of these states are derived in appendix \ref{app1}. Their action on the normalized basis states denoted by $|~\rangle$ are given by:
\begin{enumerate} 
	\item $a_1\cdot \tilde{a}_2$
	\begin{align}
		(a_1\cdot \tilde{a}_2)|~\rangle&= \alpha |s_{12}-1\rangle\nonumber\\
		\alpha&= \sqrt{s_{12}(s_{12} +r_1+r_2+1)}
	\label{su2fa1ta2}
\end{align}
\item $a^\dagger_i \cdot a_j $ 
\begin{align}
(a^\dagger_i \cdot a_j)|~\rangle &= \beta |r_j-1,r_i+1\rangle \nonumber\\
&\beta=r_j
\label{su2faiaj}
\end{align}
\item $q^\dagger \cdot a_i$ 
\begin{align}
(q^\dagger \cdot a_i) |~\rangle &=\gamma|r_i-1,v+1\rangle\nonumber\\
\gamma&=r_i\frac{[(-1)^{i}s_{12}+2]}{\sqrt{2(s_{12}+2)}}
\label{su2fqdai}
\end{align}
In the above expression, we keep in mind that when $r_i=1$, $v$ has to be zero by (\ref{su2cond}), and suppress a factor of $(1-v)$ in $\gamma$. In what follows we will freely suppress such factors for simplicity of notation whenever there is no confusion.
\item $a^\dagger_i \cdot q$
\begin{align}
(a^\dagger_i \cdot q) |~\rangle=  \delta_1 | s_{12}+1,r_j-1\rangle + \delta_2|v-1,r_i+1\rangle \hspace{1cm} ; i\neq j \nonumber\\
\delta_1=[(-1)^i r_j] \sqrt{{s_{12}+1}} \hspace{2cm} \delta_2=v\sqrt{{s_{12}+2}}
\end{align}
\item $a_i\cdot \tilde{q}$
\begin{align}
	(a_i\cdot \tilde{q}) |~\rangle &= \epsilon_1 |r_i-1\rangle + \epsilon_2| s_{12}-1,v-1,r_j+1\rangle \hspace{1cm} ;i\neq j\nonumber\\
	\epsilon_1&=r_i\sqrt{{s_{12}+2}} \hspace{1cm} \epsilon_2=v\sqrt{s_{12}}
\end{align}		
\end{enumerate}
In the above, $i,j=1,2$ and  we use the notation $|~\rangle$ for a generic state $|s_{ij},r_i,v\rangle$ and only the quantum numbers that changes across the equality are mentioned in the R.H.S.

\section{SU(3)}
\label{su3}
Bosonic representation of SU(3) generators\cite{rsarx,su3rmi,mc} are given by 
\begin{align}
E^a= a^\dagger \frac{\lambda^a}{2} a - b^\dagger \frac{\lambda^{*a}}{2} b
\end{align}
where $\lambda^a$ are the eight generators of SU(3) and $a,b$ are irreducible SU(3) Schwinger bosons satisfying the following commutation \cite{su3rmi,rsarx} relations which preserve $a^\dagger \cdot b^\dagger \approx 0$, a condition which allows labelling of all irreducible adjoint representations uniquely\cite{su3rmi}.
\begin{align}
\big[a_\alpha~,~a^\dagger_\beta\big]&= \delta_{\alpha\beta}-\tilde{N}~ b^\dagger_\alpha b_\beta \nonumber \\
\big[b_\alpha~,~b^\dagger_\beta\big]&= \delta_{\alpha\beta}-\tilde{N}~ a^\dagger_\alpha a_\beta \nonumber\\
\big[a_\alpha~,~b^\dagger_\beta\big]&= -\tilde{N}~ b^\dagger_\alpha a_\beta
\label{aircom}
\end{align} 
Above, $\tilde{N}=\frac{1}{N+2}$ where $N=N_a+N_b$ is the number operator for the total number of oscillators.
\begin{figure}
	\includegraphics[]{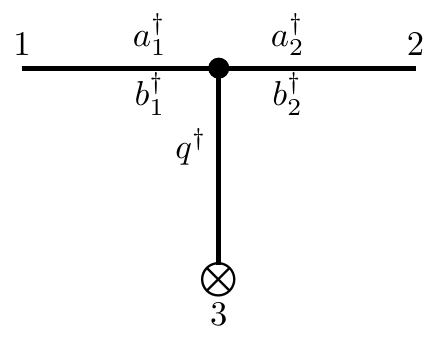}
	\caption{ Type-II vertex for SU(3). $a^\dagger_i, b^\dagger_i$ are the prepotential triplets at the vertex and $q^\dagger$ is the fermionic triplet.}
	\label{3vertexsu3}
\end{figure}
A fermionic representation of the SU(3) generator is given by: 
\begin{align}
{\cal E}^a= q^\dagger \frac{\lambda^a}{2} q
\end{align}
Where, $q_\alpha, q^\dagger_\alpha; \alpha =1,2,3$ are fermionic annihilation, creation operators satisfying the anticommutation relations $\{q_\alpha,q^\dagger_\beta\}=\delta_{\alpha\beta}, \{q_\alpha,q_\beta\}=0, \{q^\dagger_\alpha,q^\dagger_\beta\}=0$. $q^\dagger_\alpha|0\rangle$  gives a fundamental 3 representation space of SU(3).
  We are interested in constructing a singlet representation satisfying $E^a_1+E^a_2+{\cal E}^a=0$ starting with two  SU(3) irrep in the bosonic representation and one in the fermionic representation. 

A basis of such a singlet representation is given by:   
  \begin{align}
|s_{ij},r_l,t,u_i,v\rangle = \frac{1}{\sqrt{S(s_{ij},r_l,t,u_i,v)}}  \prod\limits_{\stackrel{i\neq j}{i,j=1,2}}(a_{i}^\dagger \cdot b^\dagger_{j} )^{s_{ij}} (q^\dagger \cdot b^\dagger_i)^{r_i}(\epsilon a^\dagger_1a^\dagger_2q^\dagger)^t(\epsilon a^\dagger_iq^\dagger q^\dagger)^{u_i}(\epsilon q^\dagger q^\dagger q^\dagger)^v|0\rangle 
\label{fsu3gib}
\end{align}
where, $s_{ij}\equiv s_{12},s_{21},s_{13},s_{31},s_{23},s_{32}$; $\epsilon a^\dagger_1a^\dagger_2q^\dagger=\epsilon_{\alpha \beta \gamma} a^\dagger_{1,\alpha} a^\dagger_{2,\beta} q^\dagger_\gamma$
  and the normalization factor $S(s_{ij},r_l,t,u_i,v) = {}_u\langle s_{ij},r_l,t,u_i,v|s_{ij},r_l,t,u_i,v\rangle_u$ is calculated in appendix \ref{app2}, $|s_{ij},r_l,t,u_i,v\rangle_u$ being the unnormalized basis.
  The states are all not independent due to the following relations: 
  	\begin{align}
  	\big(q^\dagger\cdot b^\dagger_1\big)\big(\epsilon a^\dagger_1a^\dagger_2q^\dagger\big)&= \frac{1}{2} \big(a^\dagger_2\cdot b^\dagger_1\big)\big(\epsilon a^\dagger_1 q^\dagger q^\dagger\big)&
  	\big(q^\dagger\cdot b^\dagger_2\big)\big(\epsilon a^\dagger_1a^\dagger_2q^\dagger\big)&= -\frac{1}{2} \big(a^\dagger_1\cdot b^\dagger_2\big)\big(\epsilon a^\dagger_2 q^\dagger q^\dagger\big)\nonumber\\
  	\big(q^\dagger\cdot b^\dagger_j\big)\big(\epsilon a^\dagger_iq^\dagger q^\dagger\big)&=\frac{1}{2} \big(a^\dagger_i\cdot b^\dagger_j\big)\big(\epsilon q^\dagger q^\dagger q^\dagger\big) &
  	\big(q^\dagger\cdot b^\dagger_i\big)\big(\epsilon a^\dagger_iq^\dagger q^\dagger\big)&=0 \nonumber\\
  	\big(q^\dagger\cdot b^\dagger_i\big)\big(\epsilon q^\dagger q^\dagger q^\dagger\big)&=0 &
  		\big(\epsilon a^\dagger_i q^\dagger q^\dagger\big)\big(\epsilon q^\dagger q^\dagger q^\dagger\big)&=0 \nonumber\\
  			\big(\epsilon a^\dagger_1 a^\dagger_2 q^\dagger\big)\big(\epsilon q^\dagger q^\dagger q^\dagger\big)&=0 &
  			\big(\epsilon a^\dagger_i q^\dagger q^\dagger\big)\big(\epsilon a^\dagger_j q^\dagger q^\dagger\big)&=0 &
  	\label{qrels}
  	\end{align}
  Therefore, a unique and complete labelling of states is achieved if the quantum numbers satisfy the following relation 
  \begin{align}
  (r_1+r_2+t+u_1+u_2)v+(r_1+r_2+t+u_1)u_2+(r_1+r_2+t)u_1+(r_1+r_2)t=0.
  \label{su3cond}
  \end{align} 
   This means that from the set $(r_1,r_2,t,u_1,u_2,v)$ no pair except $r_1,r_2$ can be simultaneously non-zero.
The action of a gauge invariant operator $O$ on a general normalised basis state is 
\begin{align}
O~|s_{ij},r_l,t,u_i,v\rangle = \sum\limits_{s'_{ij},r'_i,t',u'_i,v'} ~\Bigg[\sqrt{\frac{S(s'_{ij},r'_i,t',u'_i,v')}{S(s_{ij},r_i,t,u_i,v)}}~\Bigg]~C_{s'_{ij},r'_i,t',u'_i,v'}~|s'_{ij},r'_i,t',u'_i,v'\rangle
\end{align}
Therefore, in order to compute the action of various operators on normalized states, one only need ratio of norms and the action of those operators on the unnormalized states. These are calculated in appendix \ref{app2}. Action of various invariant operators on the normalized basis states are listed below. 
\begin{figure}
	\centering
	\includegraphics[scale=1.3]{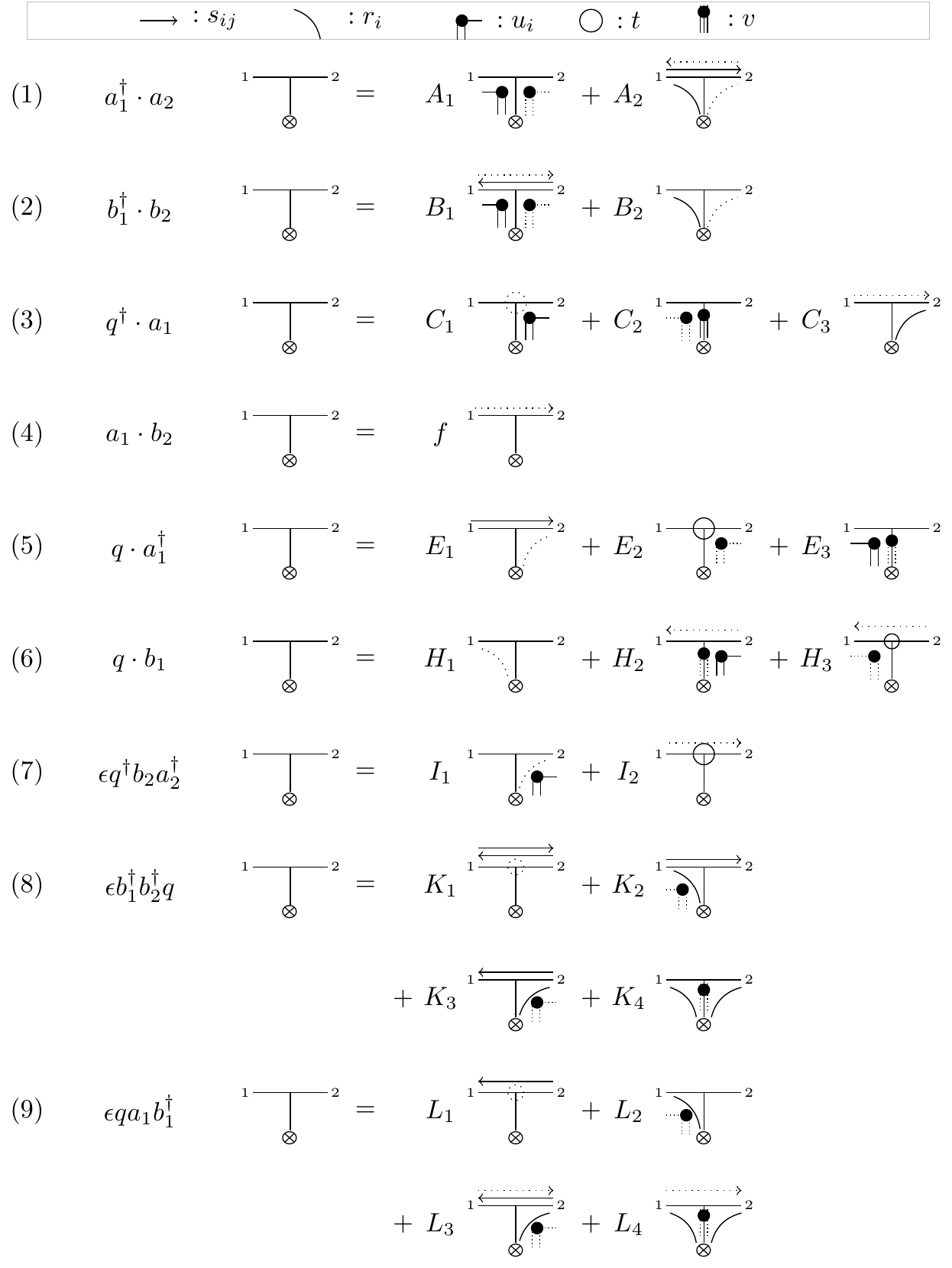}
	\caption{graphical representation of action of various SU(3) invariant operators on the basis states. Increase/decrease in any quantum number is denoted with a solid/dotted line.}
	\label{fsu3opaction}
\end{figure}

\begin{figure}
	\centering
	\includegraphics[scale=1.3]{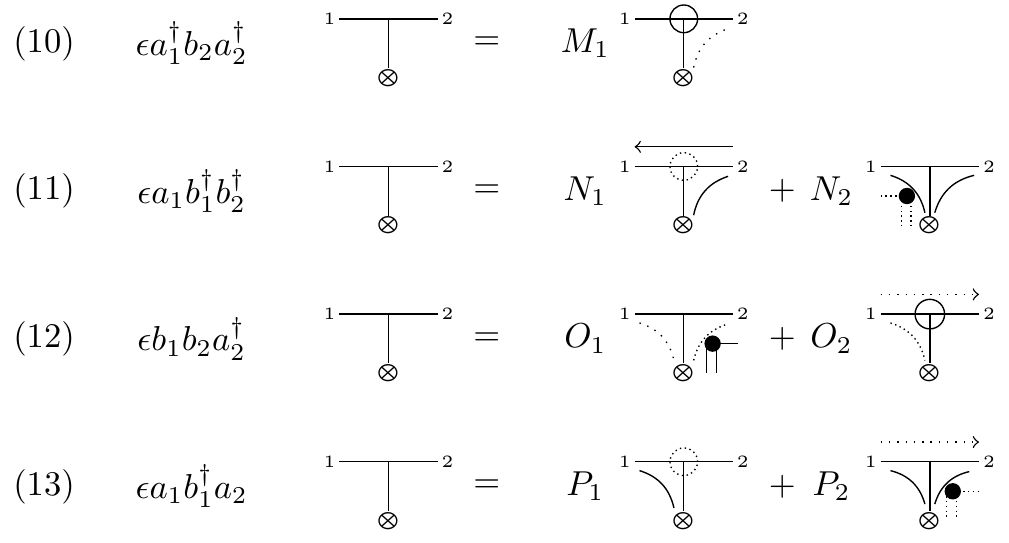}
	\caption{graphical representation of action of various SU(3) invariant operators containing three bosonic operators on the basis states.}
	\label{fsu3opaction2}
\end{figure}

\begin{enumerate}
	\item $a^\dagger_1\cdot a_2$
	{\footnotesize
		\begin{align}
		&(a^\dagger_1\cdot a_2)|
		~\rangle= A_1 \Bigg|\begin{matrix}u_2-1\\u_1+1\end{matrix}\Bigg\rangle+A_2\Bigg|\begin{matrix}s_{12}+1,s_{21}-1\\r_2-1,r_1+1\end{matrix}\Bigg\rangle \nonumber\\
		&A_1=\bigg(\frac{s_{21}+2}{s_{12}+s_{21}+2}\bigg)u_2  \hspace{1cm} A_2= 
		\bigg(\frac{-s_{21}}{s_{21}+s_{12}+2}\bigg)(1-r_1)r_2 \sqrt{\frac{\bar{f}^{12}(s_{12}+1)}{\bar{f}^{21}(s_{12}=0,s_{21})}}
		\label{su3fad1a2}
		\end{align}}
	\item $b^\dagger_1\cdot b_2$
	{\footnotesize 
		\begin{align}
		&(b^\dagger_1\cdot b_2)|~\rangle=B_1\Bigg|\begin{matrix}s_{21}+1,s_{12}-1\\u_2-1,u_1+1\end{matrix}\Bigg\rangle+ B_2\Bigg|\begin{matrix}r_2-1\\r_1+1\end{matrix}\Bigg\rangle \nonumber\\
		& B_1=u_2\bigg(\frac{-s_{12}}{s_{12}+s_{21}+2}\bigg) \sqrt{\frac{\bar{f}^{21}(s_{12}=0,s_{21}+1)}{\bar{f}^{12}(s_{12})}}\hspace{1cm} B_2= \bigg(\frac{s_{12}+2}{s_{12}+s_{21}+2}\bigg)(1-r_1)r_2
		\label{bd1b2}
		\end{align}}
	\item $q^\dagger \cdot a_i$ 
		{\footnotesize\begin{align}
		&(q^\dagger \cdot a_1)~|~\rangle= C_1\Bigg| \begin{matrix}
		t-1\\u_2+1
		\end{matrix}\Bigg\rangle+C_2\Bigg|\begin{matrix}
		u_1-1\\v+1
		\end{matrix}\Bigg\rangle+ C_3 \Bigg|\begin{matrix}
		s_{12}-1\\r_2+1
		\end{matrix}\Bigg\rangle\nonumber\\
		& C_1= \sqrt{2}t\hspace{1cm} C_2=\sqrt{3}u_1  \hspace{1cm} C_3= (1-r_2)s_{12}\Big(1-\frac{r_1}{s_{12}+s_{21}+2}\Big)\sqrt{\frac{1}{\bar{f}^{12}(s_{12})}}\sqrt{\frac{3(r_1+1)}{1+2r_1}} 
		\label{fsu3qda1}
		\end{align}}

	{\footnotesize
		\begin{align}
		&(q^\dagger \cdot a_2)~|~\rangle=D_1\Bigg|\begin{matrix}
		t-1\\u_1+1
		\end{matrix}\Bigg\rangle+ D_2 \Bigg|\begin{matrix}
		u_2-1\\v+1
		\end{matrix}\Bigg\rangle+ D_3 \Bigg|\begin{matrix}
		s_{21}-1\\r_1+1
		\end{matrix}\Bigg\rangle \nonumber\\
		& D_1= 2t\hspace{1cm} D_2=3u_2  \hspace{1cm} D_3= (1-r_1)s_{21}\Big(1-\frac{r_2}{s_{21}+s_{12}+2}\Big)\sqrt{\frac{1}{\bar{f}^{21}(s_{12}=0,s_{21})}}\sqrt{\frac{3(r_2+1)}{1+2r_2}} 
		\label{qda2}
		\end{align}}

	\item 
	$a_1\cdot b_2$.

	\begin{align}
	&(a_1\cdot b_2)|~\rangle_u=f|s_{12}-1\rangle  \nonumber\\
	&f= \sqrt{{\bar{f}^{12}(s_{12})}}
	\end{align}
	\item $q\cdot a^\dagger_i$ 
	\begin{align} 
	&(q.a^\dagger_1)|~ \rangle= e_1\Bigg|\begin{matrix}
	s_{12}+1\\r_2-1
	\end{matrix}\Bigg\rangle+e_2\Bigg|\begin{matrix}
	t+1\\u_2-1
	\end{matrix}\Bigg\rangle+e_3\Bigg|\begin{matrix}
	u_{1}+1\\v-1
	\end{matrix}\Bigg\rangle \nonumber\\
	& e_1=r_2\sqrt{\frac{1+2r_1}{3(r_1+1)}} \sqrt{\bar{f}^{12}(s_{12}+1)} \hspace{1cm} e_2= - (1-r_1)(1-r_2)u_2\sqrt{2}\nonumber\\
	&\hspace{2.5cm} e_3=(1-r_1)(1-r_2)v\sqrt{3}
	\label{su3fqad1}
	\end{align}
	\begin{align}
	&(q.a^\dagger_2)|~ \rangle= g_1\Bigg|\begin{matrix}
	s_{21}+1\\r_1-1
	\end{matrix}\Bigg\rangle+ g_2\Bigg|\begin{matrix}
	t+1\\u_1-1
	\end{matrix}\Bigg\rangle+g_3\Bigg|\begin{matrix}
	u_{2}+1\\v-1
	\end{matrix}\Bigg\rangle \nonumber\\
	&g_1 =r_1\sqrt{\frac{1+2r_2}{3(1+r_2)}}\sqrt{\bar{f}^{21}(s_{12}=0,s_{21}+1)} \hspace{1cm} g_2=(1-r_1)(1-r_2)u_1\sqrt{2}\nonumber\\
	& \hspace{2.5cm} g_3= (1-r_1)(1-r_2)v\sqrt{3}
	\label{qad2}
	\end{align}
	\item $q\cdot b_i$
	\begin{align}
	(q\cdot b_1) |~ \rangle=&H_1|r_1-1\rangle+H_2\Bigg|\begin{matrix}
	s_{21}-1,v-1\\u_2+1
	\end{matrix}\Bigg\rangle+H_3\Bigg|\begin{matrix}s_{21}-1,u_1-1\\t+1\end{matrix}\Bigg\rangle\nonumber\\
	&H_1=(s_{21}+\frac{s_{12}(1-r_2)}{s_{12}+s_{21}+2}-3)r_1 \sqrt{\frac{(1+2r_2)}{3(1+r_2)}} \hspace{1cm} H_2=\sqrt{3} s_{21}v\sqrt{\frac{1}{\bar{f}^{21}(s_{12}=0,s_{21})}}\hspace{1cm} \nonumber\\
	&H_3= \sqrt{2}u_1s_{21}\Big(1+\frac{1}{s_{12}+s_{21}+2}\Big) \sqrt{\frac{1}{\bar{f}^{21}(s_{12}=0,s_{21})}}
	\end{align}
	\item $\epsilon q^\dagger b_2a^\dagger_2$
	\begin{align}
	&(\epsilon q^\dagger b_2 a^\dagger_2) |~\rangle =I_1\Bigg|\begin{matrix}r_2-1\\u_2+1\end{matrix}\Bigg\rangle+ I_2 \Bigg|\begin{matrix}s_{12}-1\\t+1\end{matrix}\Bigg\rangle \nonumber\\
	& I_1=2r_2\sqrt{\frac{(1+2r_1)}{1+r_1}} \hspace{1cm} I_2= \frac{\sqrt{6}(1-t)(1-u_i)(1-r_i)(1-v) s_{12}}{\sqrt{\bar{f}^{12}(s_{12})}}
	\end{align}
	\begin{align}
	&(\epsilon q^\dagger b_1 a^\dagger_2) |~\rangle =J_1 \Bigg|\begin{matrix}r_1-1\\u_2+1\end{matrix}\Bigg\rangle\nonumber\\
	&J_1=2r_1\frac{(1+2r_2)}{(1+r_2)}
	\end{align} 
	\item  $\epsilon b^\dagger_1b^\dagger_2 q$
	{\footnotesize
		\begin{align}
		&(\epsilon b^\dagger_1b^\dagger_2q)|~\rangle =K_1 \Bigg|\begin{matrix}s_{12}+1,s_{21}+1\\t-1\end{matrix}\Bigg\rangle+K_2\Bigg|\begin{matrix}r_1+1,s_{12}+1\\u_1-1\end{matrix}\Bigg\rangle +K_3\Bigg|\begin{matrix}r_2+1,s_{21}+1\\u_2-1\end{matrix}\Bigg\rangle+K_4\Bigg|\begin{matrix}
		r_1+1,r_2+1\\v-1
		\end{matrix}\Bigg\rangle \nonumber\\
		&K_1=-\frac{t}{\sqrt{6}}\sqrt{\bar{f}^{12}(s_{12}+1)\bar{f}^{21}(s_{12}=0,s_{21}+1)} \hspace{1cm} K_2=(1-r_1)(1-r_2) {u_1}\sqrt{\bar{f}^{12}(s_{12}+1)}\nonumber\\
		&K_3={u_2}\sqrt{\bar{f}^{21}(s_{12}=0,s_{21}+1)} \hspace{1cm} K_4= v(1-r_1)(1-r_2)
		\end{align}}
	\item  $\epsilon q a_1b^\dagger_1$

	{\footnotesize
		\begin{align}
		&(\epsilon q a_1 b^\dagger_1)|~\rangle=L_1\Bigg|\begin{matrix}
		s_{21}+1\\t-1
		\end{matrix}\Bigg\rangle+L_2\Bigg|\begin{matrix}
		r_1+1\\u_1-1
		\end{matrix}\Bigg\rangle+L_3\Bigg|\begin{matrix}
		s_{12}-1,s_{21}+1\\u_2-1,r_2+1
		\end{matrix}\Bigg\rangle+L_4\Bigg|\begin{matrix}
		s_{12}-1,r_1+1\\v-1,r_2+1
		\end{matrix}\Bigg\rangle\nonumber\\
		&L_1=(s_{12}+4)\frac{t}{\sqrt{6}}\sqrt{\bar{f}^{21}(s_{12}=0,s_{21}+1)} \hspace{1cm} L_2= (s_{12}+2){u_1} \nonumber\\
		& L_3= {-s_{12}u_2} \sqrt{\frac{\bar{f}^{21}(s_{12}=0,s_{12}+1)}{\bar{f}^{12}(s_{12})}} \hspace{1cm} L_4= -\sqrt{6} s_{12} v \sqrt{\frac{1}{\bar{f}^{12}(s_{12})}}
		\end{align}}
\item $\epsilon a^\dagger_1b_2a^\dagger_2$
\begin{align}
&(\epsilon a^\dagger_1b_2a^\dagger_2) |\rangle =  M_1\bigg|\begin{matrix}
r_2-1\\t+1
\end{matrix}\bigg\rangle\nonumber\\
& M_1=-\bigg(\frac{s_{12}+s_{21}+4}{s_{12}+s_{21}+3}\bigg)r_{2}(1-r_1)\sqrt{2}
\end{align}
\item $\epsilon a_1b^\dagger_1b^\dagger_2$
\begin{align}
&(\epsilon a_1b^\dagger_1b^\dagger_2)|\rangle= N_1\bigg|\begin{matrix}
t-1,r_2+1\\s_{21}+1
\end{matrix}\bigg\rangle +N_2\bigg|\begin{matrix}
r_1+1,r_2+1\\u_1-1
\end{matrix}\bigg\rangle\nonumber\\
&N_1=\bigg(\frac{s_{12}+s_{21}+4}{s_{12}+s_{21}+3}\bigg)t(1-r_2)\sqrt{\frac{\bar{f}^{12}(s_{12}=0,s_{21}+1)(r_1+1)}{2(1+2r_1)}} \nonumber\\
& N_2=\bigg(\frac{s_{12}+s_{21}+4}{s_{12}+s_{21}+3}\bigg)\frac{u_1}{\sqrt{2}}
\end{align}
\item $\epsilon b_1b_2a^\dagger_2$

\begin{align}
&(b_1b_2a^\dagger_2)|\rangle=  O_1\bigg|\begin{matrix}
r_1-1,r_2-1\\u_2+1
\end{matrix}\bigg\rangle+O_2\bigg|\begin{matrix}
s_{12}-1,t+1\\r_1-1
\end{matrix}\bigg\rangle\nonumber\\
&O_1=\bigg(\frac{s_{12}+s_{21}+3}{s_{12}+s_{21}+2}\bigg)^2 \frac{2}{3}r_1r_2\sqrt{2}\nonumber\\
&O_2=\bigg(\frac{s_{12}+s_{21}+3}{s_{12}+s_{21}+2}\bigg)^2(-s_{12}r_1(1-r_2))\sqrt{\frac{2(1+2r_2)}{\bar{f}^{12}(s_{12})(1+r_2)}}
\end{align}
\item $\epsilon a_1b^\dagger_1a_2$
\begin{align}
&(\epsilon a_1b^\dagger_1a_2)|\rangle=P_1 \bigg|\begin{matrix}
t-1\\r_1+1\end{matrix}\bigg\rangle+P_2\bigg|\begin{matrix}u_2-1,s_{12}-1\\r_1+1,r_2+1
\end{matrix}\bigg\rangle\nonumber\\
& P_1=-\bigg(\frac{s_{12}+s_{21}+4}{s_{12}+s_{21}+2}\bigg)\bigg[\frac{(s_{12}+2)s_{21}}{(s_{12}+s_{21}+3)(s_{12}+3)}+\frac{2}{3}-\frac{s_{12}}{3(s_{12}+3)}+s_{12}\bigg]t \sqrt{\frac{r_2+1}{2(1+2r_2)}}\nonumber\\
&P_2=-\bigg(\frac{s_{12}+s_{21}+3}{s_{12}+s_{21}+2}\bigg)^2s_{12}u_2 \sqrt{\frac{1}{2\bar{f}^{12}(s_{12})}}
\end{align}
\end{enumerate}
This is also pictorially shown in fig. \ref{fsu3opaction}. Change in different quantum numbers are denoted by various symbols as tabulated at the top of the figure. Solid symbols denote an increase in the quantum number by 1 and dotted symbol denotes a decrease by 1. We remark that if the fermion $q^\dagger$ is in the $3^*$ representation, essentially the same results follow with $a^\dagger$ interchanged with $b^\dagger$ in the above expressions.
 
\section{Concluding remarks.}
\label{con}
Point splitting technique reduces Gauss law constraint in lattice gauge theory in any dimension into the problem discussed in this paper.
The construction described here, helps us to describe SU(3) lattice gauge theory within its physical Hilbert space without any redundant gauge degrees of freedom. Study of lattice QCD dynamics within such a description will be discussed in a future work.

The type-II vertex one gets on inclusion of scalar matter fields is essentially a subset of the already solved \cite{rsarx} type-I vertex with either the number of $b$ or number of $a$ being zero.   
In order to construct higher SU(3) irreps using fermionic operators, one requires more independent fermions. Such cases could involve, 
two type of fermions in a type-II 3-vertex and one can construct its singlet space in essentially the same way as described in this paper. Since the point splitting scheme makes such a construction unnecessary in lattice QCD, we don't discuss it here. Our deductions can be generalized to SU(N) using SU(N) Schwinger bosons\cite{sunb} and fermions.

\noindent{\it acknowledgement:
TPS would like to thank Manu Mathur for discussions and hospitality at S N Bose National Centre for Basic Sciences where part of the work was completed.} 

\appendix
\section{SU(2)}
\label{app1}
\subsection{Action of invariant operators.}
The action of various invariant operators are computed below: 
\begin{enumerate}
	\item $a_1 \cdot \tilde{a}_2$ 
	\begin{align}
		&(a_1 \cdot \tilde{a}_2)|s_{12},r_i,v\rangle_u= (a_1 \cdot \tilde{a}_2)(a^\dagger _1\cdot \tilde{a}^\dagger_2)|s_{12}-1\rangle_u= (2+a^\dagger_2\cdot a_2 +a^\dagger_1\cdot a_1+(a^\dagger_1\cdot \tilde{a}^\dagger_2)(a_1\cdot\tilde{a}_2)) | s_{12}-1\rangle_u\nonumber\\
	&=\sum\limits_{r=0}^{s_{12}-1} (a_1^\dagger \cdot \tilde{a}^\dagger_2)^r (2+N_1+N_2)|s_{12}-1-r\rangle_u + (a^\dagger_1\cdot\tilde{a}^\dagger_2)^{s_{12}} (a_1\cdot \tilde{a}_2)|s_{12}=0,r_i\rangle_u \nonumber\\
	&=\sum\limits_{r=0}^{s_{12}-1}(2(s_{12}-r)+r_1+r_2) |s_{12}-1\rangle_u\nonumber\\
		&= s_{12}(s_{12} +r_1+r_2+1) |s_{12}-1\rangle_u
				\label{ua1a2}
		\end{align}
	In the third step we have used the fact that only one among $r_1,r_2,v$ can be non-zero.
	\item $a^\dagger_i \cdot a_j $ 
	\begin{align}
	(a^\dagger_i \cdot a_j)|s_{12},r_i,v\rangle_u = (a^\dagger_1\cdot \tilde{a}^\dagger_2)^{s_{12}}(r_j)(a^\dagger_i\cdot a_j)(a^\dagger_j \cdot \tilde{q}^\dagger)|s_{12}=0,r_j-1\rangle_u = r_j | s_{12},r_j-1,r_i+1,v\rangle_u 
	\end{align}
	\item $q^\dagger \cdot a_i$ 
	\begin{align}
	(q^\dagger \cdot a_1) |s_{12},r_i,v\rangle_u &= (q^\dagger \cdot a_1)(a^\dagger_1 \cdot \tilde{a}^\dagger_2) |s_{12}-1,r_i,v\rangle_u\nonumber\\
	&= [-(a^\dagger_2 \cdot \tilde{q}^\dagger)+(a^\dagger_1 \cdot \tilde{a}^\dagger_2)(q^\dagger \cdot a_1)]|s_{12}-1\rangle \nonumber\\
	&=r_1(-\frac{1}{2}s_{12}+1)|r_1-1,v+1\rangle_u
	\end{align}
	We have used the relation $(a^\dagger_1 \cdot \tilde{q}^\dagger)(a^\dagger_2\cdot \tilde{q}^\dagger)=\frac{1}{2} (a^\dagger_1\cdot \tilde{a}^\dagger_2)(q^\dagger \cdot \tilde{q}^\dagger) $
	Similarly, 
	\begin{align}
	(q^\dagger \cdot a_2) |s_{12},r_i,v\rangle_u = r_2(\frac{1}{2}s_{12}+1)|r_2-1,v+1\rangle_u
	\end{align}
	The difference in sign in the above expression is due to our convention of choosing $a^\dagger_1\cdot \tilde{a}^\dagger_2$ instead of $a^\dagger_2\cdot \tilde{a}^\dagger_1$ when defining the states. 
	\item $a^\dagger_i \cdot q$
	\begin{align}
	(a^\dagger_1 \cdot q) |s_{12},r_i,v\rangle_u=  -r_2 | s_{12}+1,r_2-1\rangle_u + 2v|v-1,r_1+1\rangle_u 
	\end{align}
	We have used $(a^\dagger_1 \cdot q)(\tilde{a}^\dagger_1\cdot q^\dagger)|0\rangle=0$, $(a^\dagger_1 \cdot q)(a^\dagger_2\cdot \tilde{q}^\dagger)|0\rangle=-(a^\dagger_1\cdot \tilde{a}^\dagger_2)|0\rangle$, $(a^\dagger\cdot q)(q^\dagger\cdot \tilde{q}^\dagger)|0\rangle= 2(a^\dagger_1\cdot \tilde{q}^\dagger)|0\rangle$. Similarly, 
	\begin{align}
	(a^\dagger_2 \cdot q) |s_{12},r_i,v\rangle_u=  r_1 | s_{12}+1,r_1-1\rangle_u + 2v|v-1,r_2+1\rangle_u 
	\end{align}
	\item $a_i\cdot \tilde{q}$
	\begin{align}
&	(a_1\cdot \tilde{q}) |s_{12},r_i,v\rangle_u = (a_1\cdot \tilde{q})(a^\dagger_1 \cdot \tilde{a}^\dagger_2) |s_{12}-1,r_i,v\rangle_u =[(q\cdot a^\dagger_2)+(a^\dagger_1\cdot \tilde{a}^\dagger_2)(a_1\cdot \tilde{q})]|s_{12}-1,r_i,v\rangle_u \nonumber\\
	&=s_{12}(q\cdot a^\dagger_2)|s_{12}-1,r_i,v\rangle+(a^\dagger \cdot \tilde{a}^\dagger_2)^{s_{12}}(a_1\cdot \tilde{q})|s_{12}=0,r_i,v\rangle_u\nonumber\\
	&= (s_{12}+2)r_1 |r_1-1\rangle_u +s_{12}2v | s_{12}-1,v-1,r_2+1\rangle_u
	\end{align}
	In the last step, we have used the relation $(a_1\cdot \tilde{q})(a^\dagger_1\cdot \tilde{q}^\dagger)=2-q^\dagger \cdot q$. Similarly, 
{\footnotesize
	\begin{align}
	(a_2\cdot \tilde{q}) |s_{12},r_i,v\rangle_u =(s_{12}+2)r_2 |r_2-1\rangle_u +s_{12}(2v) | s_{12}-1,v-1,r_1+1\rangle_u
	\end{align}}
\end{enumerate}
\subsection{Norm}
The norm of $|s_{12},r_i,v \rangle_u$ is computed as below. 
\begin{align}
	{}_u\langle s_{12},r_i,v|s_{12},r_i,v\rangle_u&={}_u\langle s_{12}-1,r_i|a_1\cdot \tilde{a}_2|s_{12},r_i\rangle_u\nonumber\\
	&=s_{12}(s_{12}+r_1+r_2+1) ~{}_u\langle s_{12}-1|s_{12}-1\rangle_u
	\end{align}
The above relation can be iterated to give 
\begin{align}
	{}_u\langle s_{12},r_i,v|s_{12},r_i,v\rangle_u = \frac{s_{12}!(s_{12}+r_1+r_2+1)!}{(r_1+r_2+1)!} ~{}_u\langle s_{12}=0,r_i,v|s_{12}=0,r_i,v\rangle_u
\end{align}
Since
\begin{align} 
\langle0| (a_i\cdot \tilde{q})(a^\dagger_i\cdot \tilde{q}^\dagger)|0\rangle=\langle0|(2+a^\dagger_i \cdot a_i-q^\dagger\cdot q-(a^\dagger_i\cdot \tilde{q})(a_i\cdot \tilde{q}))|0\rangle =2 
\end{align}
and 
\begin{align}
\langle0|(\tilde{q}\cdot q)(q^\dagger \cdot \tilde{q}^\dagger)|0\rangle=4 \langle0|(q_2q_1)(q^\dagger_1q^\dagger_2)|0\rangle=4
\end{align}
we have, 
\begin{align}
&{}_u\langle s_{12}=0,r_1,r_2,v|s_{12}=0,r_1,r_2,v \rangle_u= (r_1+r_2+3v+1)
\end{align}
The norm is given by 
\begin{align}
{}_u\langle s_{12},r_i,v|s_{12},r_i,v\rangle_u  = \frac{s_{12}!(s_{12}+r_1+r_2+1)!}{(r_1+r_2+1)!}(r_1+r_2+3v+1)
\label{su2norm1}
\end{align}

\section{SU(3)}
\label{app2}
\subsection{Action of invariant operators}
The action of a gauge invariant operator $O$ on the $|s_{ij},r_i,t,u_i,v\rangle_u$ can be computed by using the commutation relations (\ref{aircom}) to shift $O$ to the right until it hits $|0\rangle$.	In the following, whenever there is no confusion, only the $s_{ij},r_i,t,u_i,v$ values which changes across an equation are written. Also, it is convenient to introduce the following notation $(i\cdot j)\equiv a^\dagger_i\cdot b^\dagger_j$; $|~\rangle \equiv |s_{ij},r_i,t,u_i,v\rangle$ .

\begin{enumerate}
	\item $a^\dagger_1\cdot a_2$
	\begin{align}
	(a^\dagger_1\cdot a_2) |~\rangle_u = (a^\dagger_1\cdot a_2) (a^\dagger_1\cdot b^\dagger_2) |s_{12}-1\rangle_u
	\end{align}
	
	Using (\ref{aircom}), 
	\begin{align}
	(a^\dagger_1\cdot a_2) (a^\dagger_1\cdot b^\dagger_2\Big)
	&=(1-\tilde{N}_2)(a^\dagger_1\cdot b^\dagger_2)(a^\dagger_1\cdot a_2)
	\end{align}
	Repeating this $s_{12}$ times we get, 
	\begin{align}
	(a^\dagger_1\cdot a_2)|
	~\rangle_u&=\Big[(1-\tilde{N}_1)(1\cdot 2)\Big]^{s_{12}}(a^\dagger_1 \cdot a_2)|s_{12}=0\rangle_u\nonumber\\
	&=(1\cdot2)^{s_{12}}\left(\frac{\hat{N}_2+2}{\hat{N}_2+s_{12}+2}\right)(a^\dagger_1\cdot a_2)|s_{12}=0\rangle_u
	\label{I1it}
	\end{align}
	Using the relation 
	$	(a^\dagger_1\cdot a_2) (a^\dagger_2\cdot b^\dagger_1)=(a^\dagger_2\cdot b^\dagger_1)(a^\dagger_1\cdot a_2) -\tilde{N}_2(a^\dagger_1\cdot b^\dagger_2)( b^\dagger_1\cdot b_2)$
	repeatedly, we get 
	\begin{align}
	(a^\dagger_1\cdot a_2) |s_{12}=0\rangle_u &= (a^\dagger _2 \cdot b^\dagger_1)^{s_{21}} (a^\dagger_1\cdot a_2)|s_{12}=0,s_{21}=0\rangle_u\nonumber\\ 
	&+ \sum\limits_{r=0}^{s_{21}-1}(a^\dagger_2 \cdot b^\dagger_1)^r \bigg\{-\tilde{N}_2(a^\dagger_1\cdot b^\dagger_2)( b^\dagger_1\cdot b_2) \bigg\}|s_{12}=0,s_{21}-1-r\rangle_u
	\label{Iit2pre}
	\end{align}
	Using, $(b^\dagger_1\cdot b_2)(2\cdot 1)= (1-\tilde{N}_2)(2\cdot1)(b^\dagger_1\cdot b_2)$, we get 
	\begin{align}
	( b^\dagger_1\cdot b_2) |s_{12}=0,s_{21}-1-r\rangle_u=(2\cdot 1)^{s_{21}-1-r} \bigg(\frac{\hat{N}_2+2}{\hat{N}_2+s_{21}+1-r}\bigg)(b^\dagger_1\cdot b_2)|s_{12}=0,s_{21}=0\rangle_u
	\end{align}
	Therefore, 
	{\footnotesize
	\begin{align}
&(a^\dagger_1\cdot a_2)|
~\rangle_u=(1\cdot2)^{s_{12}}(2\cdot 1)^{s_{21}}\left(\frac{\hat{N}_2+s_{21}+2}{\hat{N}_2+s_{12}+s_{21}+2}\right)(a^\dagger_1\cdot a_2)|s_{ij}=0\rangle_u \nonumber\\
&+\frac{(\hat{N}_2-s_{12}+2)(\hat{N}_2-s_{12}-s_{21}
	+2)}{\hat{N}_2+2} \sum\limits_{r=0}^{s_{21}-1} \bigg\{\frac{-1}{(\hat{N}_2+s_{21}-r+2)(\hat{N}_2+s_{21}-r+1)}\bigg\}(b^\dagger_1 \cdot b_2) \bigg|\begin{matrix}
	s_{12}+1\\s_{21}-1
\end{matrix}\bigg\rangle_u
\end{align}}
Now, 
\begin{align}
(a^\dagger_1\cdot a_2) |s_{ij}=0\rangle_u &= u_2 |u_2-1,u_1+1\rangle_u \nonumber\\
(b^\dagger_1 \cdot b_2)|s_{ij}=0 \rangle_u&= (1-r_1)r_2 |r_2-1,r_1+1\rangle_u
\end{align}
 Therefore, 
 {\footnotesize
 \begin{align}
 &(a^\dagger_1\cdot a_2)|
~\rangle_u= \bigg(\frac{s_{21}+2}{s_{12}+s_{21}+2}\bigg)u_2 \Bigg|\begin{matrix}u_2-1\\u_1+1\end{matrix}\Bigg\rangle_u+\bigg(\frac{-s_{21}}{(s_{21}+s_{12}+2)}\bigg)(1-r_1)r_2\Bigg|\begin{matrix}s_{12}+1,s_{21}-1\\r_2-1,r_1+1\end{matrix}\Bigg\rangle_u
 \label{ad1a2}
 \end{align}}
 \item $b^\dagger_1\cdot b_2$
 Similarly, 
 {\footnotesize 
 \begin{align}
 (b^\dagger_1\cdot b_2)|~\rangle_u=u_2\bigg(\frac{-s_{12}}{s_{12}+s_{21}+2}\bigg)\Bigg|\begin{matrix}s_{21}+1,s_{12}-1\\u_2-1,u_1+1\end{matrix}\Bigg\rangle_u+ \bigg(\frac{s_{12}+2}{s_{12}+s_{21}+2}\bigg)(1-r_1)r_2\Bigg|\begin{matrix}r_2-1\\r_1+1\end{matrix}\Bigg\rangle_u
 \label{bd1b2}
 \end{align}}
\item $q^\dagger \cdot a_i$ 
\begin{align}
&(q^\dagger \cdot a_2)~|~\rangle_u= \bigg[(1\cdot 2)\Big(q^\dagger\cdot a_2\Big)- \tilde{N}_2 \Big(q^\dagger\cdot b^\dagger_2\Big)\Big(a^\dagger_1\cdot a_2\Big)\bigg] |s_{12}-1\rangle_u \nonumber \\
&= (1\cdot 2)^{s_{12}}\Big(q^\dagger \cdot a_2\Big) \Big|s_{12}=0\rangle_u + \sum\limits_{r=0}^{s_{12}-1}(1\cdot 2)^r \bigg\{-\tilde{N}_2 \Big(q^\dagger\cdot b^\dagger_2\Big)\Big(a^\dagger_1\cdot a_2\Big)\bigg\} |s_{12}-1-r\rangle_u \nonumber\\
\end{align}
Now,
\begin{align}
&(q^\dagger \cdot a_2)|s_{12}=0\rangle_u=\bigg\{(2\cdot 1) \Big(q^\dagger\cdot a_2\Big)+q^\dagger \cdot b^\dagger_1-\tilde{N}_2 \Big(q^\dagger\cdot b^\dagger_2\Big)\Big(b_2\cdot b^\dagger_1\Big)\bigg\}\Bigg|\begin{matrix}
s_{12}=0\\s_{21}-1
\end{matrix}\Bigg\rangle_u \nonumber\\
&=(2\cdot 1)^{s_{21}}(q^\dagger \cdot a_2) |s_{ij}=0\rangle_u +\sum\limits_{r=0}^{s_{21}-1}(2\cdot 1)^r\bigg\{q^\dagger \cdot b^\dagger_1-\tilde{N}_2\Big(q^\dagger\cdot b^\dagger_2\Big)\Big(b_2\cdot b^\dagger_1\Big)\bigg\}\Bigg|\begin{matrix}
s_{12}=0\\s_{21}-1-r
\end{matrix}\Bigg\rangle_u 
\end{align}
Therefore, 
\begin{align}
&(q^\dagger \cdot a_2)~|~\rangle_u= (1\cdot 2)^{s_{12}}(2\cdot 1)^{s_{21}} (q^\dagger \cdot a_2) |s_{ij}=0\rangle_u \nonumber\\
&+ (1\cdot 2)^{s_{12}} \sum\limits_{r=0}^{s_{21}-1} (2 \cdot 1)^{r}\Big[ -\tilde{N}_2 (q^\dagger\cdot b^\dagger_2)(b^\dagger_1\cdot b_2)\Big]\Bigg|\begin{matrix}
s_{12}=0\\s_{21}-1-r
\end{matrix}\Bigg\rangle_u +s_{21} \bigg|\begin{matrix}s_{21}-1\\r_1+1\end{matrix} \bigg\rangle_u\nonumber\\
&+\sum\limits_{r=0}^{s_{12}-1} (1\cdot2)^r\Big[-\tilde{N}_2 (q^\dagger \cdot b^\dagger_2)(a^\dagger_1\cdot a_2)\Big]|s_{12}-1-r\rangle_u
\end{align}
Using $ (q^\dagger\cdot a_2) |s_{ij}=0\rangle_u=t|s_{ij}=0,t-1,u_1+1\rangle+ u_2|s_{ij}=0,u_{2}-1,v+1\rangle_u
$ and  (\ref{ad1a2}) and (\ref{bd1b2}), we get : 
{\footnotesize\begin{align}
&(q^\dagger \cdot a_2)~|~\rangle_u=t\Bigg|\begin{matrix}
t-1\\u_1+1
\end{matrix}\Bigg\rangle_u+ u_2 \Bigg|\begin{matrix}
u_2-1\\v+1
\end{matrix}\Bigg\rangle_u+ (1-r_1)s_{21}\Big(1-\frac{r_2}{s_{12}+s_{21}+2}\Big) \Bigg|\begin{matrix}
s_{21}-1\\r_1+1
\end{matrix}\Bigg\rangle_u
\label{qda2}
\end{align}}
Similarly, 
{\footnotesize\begin{align}
&(q^\dagger \cdot a_1)~|~\rangle_u= t\Bigg| \begin{matrix}
t-1\\u_2+1
\end{matrix}\Bigg\rangle_u+u_1 \Bigg|\begin{matrix}
u_1-1\\v+1
\end{matrix}\Bigg\rangle_u+ (1-r_2)s_{12}\Big(1-\frac{r_1}{s_{12}+s_{21}+2}\Big) \Bigg|\begin{matrix}
s_{12}-1\\r_2+1
\end{matrix}\Bigg\rangle_u
\label{qda1}
\end{align}}
\item 
$a_1\cdot b_2$. 

Using the commutation relations (\ref{aircom}) repeatedly, we get : 
\begin{align}
&(a_{1,\alpha} b_{2,\alpha})~(a_{1,\beta}^\dagger b_{2,\beta}^\dagger)= a_{1,\alpha}a^\dagger_{1,\beta}b^\dagger_{2,\beta}b_{2,\alpha}+a_1\cdot a^\dagger_1-\tilde{N}_2(a_1\cdot a^\dagger_2)(a^\dagger_1\cdot a_2)\nonumber\\
&=(a^\dagger_1\cdot b^\dagger_2)(a_1\cdot b_2)+b^\dagger_2\cdot b_2-\tilde{N}_1b^\dagger_{1,\alpha}b_{1,\beta}b^\dagger_{2,\beta}b_{2,\alpha}+a_1\cdot a^\dagger_1-\tilde{N}_2(a_1\cdot a^\dagger_2)(a^\dagger_1\cdot a_2)\nonumber\\
&=(a^\dagger_1\cdot b^\dagger_2)(a_1\cdot b_2)+b^\dagger_2\cdot b_2-\tilde{N}_1\bigg(b_{1,\beta}b^\dagger_{1,\alpha}b^\dagger_{2,\beta}b_{2,\alpha}-b^\dagger_2\cdot b_2+\tilde{N}_1a^\dagger_{1,\beta}a_{1,\alpha}b^\dagger_{2,\beta}b_{2,\alpha}\bigg)\nonumber\\
&\hspace{3cm}+a_1\cdot a^\dagger_1-\tilde{N}_2( a^\dagger_2\cdot a_1)(a^\dagger_1\cdot a_2)\nonumber
\end{align} 
Using this relation to shift $(a_1\cdot b_2)$ to the right we get, 
{\footnotesize
	\begin{align}
	&(a_1\cdot b_2)~|\rangle_u=\Big[(1-\tilde{N}_1^2)(1\cdot 2)\Big]^{s_{12}} (a_1\cdot b_2) |s_{12}=0\rangle+\sum\limits_{r=0}^{s_{12}-1}\Big[(1-\tilde{N}_1^2)(1\cdot 2)\Big]^r\nonumber\\&\bigg\{3+\hat{N}_{2b}+\hat{N}_{1a}+\tilde{N}_1(\hat{N}_{2b}-\hat{N}_{1b})-\tilde{N}_1(b^\dagger_2\cdot b_1)(b^\dagger_1\cdot b_2)-\tilde{N}_2( a^\dagger_2\cdot a_1)(a^\dagger_1\cdot a_2)\bigg\}|s_{12}-1-r\rangle_u
	\label{1it}
	\end{align}}
Using, 
\begin{align}
&(a_1\cdot b_2)(a^\dagger_2\cdot b^\dagger_1)=a_{1,\alpha}a^\dagger_{2,\beta}b_{2,\alpha}b^\dagger_{1,\beta}-\tilde{N}_2(a^\dagger_2\cdot a_1)(b^\dagger_1\cdot b_2)\nonumber\\
&=a^\dagger_{2,\beta}b_{2,\alpha}b^\dagger_{1,\beta}a_{1,\alpha}-\tilde{N}_1b^\dagger_{1,\alpha}a^\dagger_{2,\beta}b_{2,\alpha}a_{1,\beta}-\tilde{N}_2(a^\dagger_2\cdot a_1)(b^\dagger_1\cdot b_2)\nonumber\\
&=(2\cdot 1)(a_1\cdot b_2)-\tilde{N}_1\Big[(a^\dagger_2,\cdot a_1)(b^\dagger_1 \cdot b_2)+\tilde{N}_1(2\cdot 1)(a_1\cdot b_2)\Big]-\tilde{N}_2( a^\dagger_2\cdot a_1)(b^\dagger_1\cdot b_2)\nonumber
\end{align}
repeatedly we get,
\begin{align}
&(a_1\cdot b_2) |s_{12}=0,s_{21}\rangle_u=\Big[\Big(1-\tilde{N}_1^2\Big)(2\cdot 1)\Big]^{s_{21}}(a_1\cdot b_2) |s_{12}=0,s_{21}=0\rangle_u \nonumber\\
&+\sum_{r=0}^{s_{21}-1}\Big[\Big(1-\tilde{N}_1^2\Big)(2\cdot 1)\Big]^r\bigg\{-(\tilde{N}_1+\tilde{N}_2)(a^\dagger_2\cdot a_1)(b^\dagger_1\cdot b_2)\bigg\}|s_{12}=0,s_{21}-1-r\rangle_u
\label{I2pre1}
\end{align}
Since,  $(a_1\cdot b_2)|s_{12}=0,s_{21}=0\rangle=0$, putting back (\ref{I2pre1}) into (\ref{1it}) gives
{\footnotesize
	\begin{align}
	&(a_1\cdot b_2)|~\rangle_u=\nonumber\\
	&-\Big[(1-\tilde{N}_1^2)(1\cdot 2)\Big]^{s_{12}} \sum_{r=0}^{s_{21}-1}\Big[\Big(1-\tilde{N}_1^2\Big)(2\cdot 1)\Big]^r(\tilde{N}_1+\tilde{N}_2)(a^\dagger_2\cdot a_1)(b^\dagger_1\cdot b_2)|s_{12}=0,s_{21}-1-r\rangle_u\nonumber\\&+\sum\limits_{r=0}^{s_{12}-1}\Big[(1-\tilde{N}_1^2)(1\cdot 2)\Big]^r\nonumber\\&\bigg\{3+\hat{N}_{2b}+\hat{N}_{1a}+\tilde{N}_1(\hat{N}_{2b}-\hat{N}_{1b})-\tilde{N}_1(b^\dagger_2\cdot b_1)(b^\dagger_1\cdot b_2)-\tilde{N}_2( a^\dagger_2\cdot a_1)(a^\dagger_1\cdot a_2)\bigg\}|s_{12}-1-r\rangle_u\nonumber
	\end{align}}
Now, using (\ref{ad1a2}) and (\ref{bd1b2}) and summing over $r$, we get 
\begin{align}
(a_1\cdot b_2)|~\rangle_u&=\Bigg[\bigg(\frac{s_{12}+s_{21}+u_1+t+r_1+2}{s_{12}+s_{21}+u_1+t+r_1+1}\bigg)s_{12}(s_{12}+r_2+u_1+t+1)\nonumber\\
&-\frac{s_{12}\Big((s_{12}+1)u_2+(s_{12}+2)(1-r_1)r_2\Big)}{(s_{12}+s_{21}+2)(s_{12}+s_{21}+1)}\Bigg]|s_{12}-1\rangle \equiv \bar{f}^{12}(s_{12}) |s_{12}-1\rangle
\end{align}
\item $q\cdot a^\dagger_i$ 
\begin{align} 
(q.a^\dagger_1)|~ \rangle_u&= (1\cdot 2)^{s_{12}} (2\cdot 1)^{s_{21}}(-q^\dagger\cdot b^\dagger_1)^{r_1} \Big\{(q\cdot a^\dagger_1)(q^\dagger \cdot b^\dagger_2)^{r_2} \Big\}|s_{ij}=0,r_i=0\rangle_u \nonumber\\
&=(1\cdot 2)^{s_{12}} (2\cdot 1)^{s_{21}}(-q^\dagger\cdot b^\dagger_1)^{r_1} \Big\{ (1\cdot 2) +(- q^\dagger\cdot b^\dagger_2)^{r_2}(q\cdot a^\dagger_1)\Big\} ||s_{ij}=0,r_i=0\rangle_u\nonumber\\
&= r_2\Bigg|\begin{matrix}
s_{12}+1\\r_2-1
\end{matrix}\Bigg\rangle_u- (1-r_1)(1-r_2)2u_2\Bigg|\begin{matrix}
t+1\\u_2-1
\end{matrix}\Bigg\rangle_u+(1-r_1)(1-r_2)3v\Bigg|\begin{matrix}
u_{1}+1\\v-1
\end{matrix}\Bigg\rangle_u
\label{qad1}
\end{align}
where we have used the relation $(q\cdot a^\dagger_1)(\epsilon a^\dagger_2q^\dagger q^\dagger)|0\rangle=-2(\epsilon a^\dagger_1a^\dagger_2q^\dagger)|0\rangle$ and $(q\cdot a^\dagger_1)(\epsilon q^\dagger_1q^\dagger q^\dagger)|0\rangle=3(\epsilon a^\dagger_1q^\dagger q^\dagger)|0\rangle$
Similary, 
\begin{align}
(q.a^\dagger_2)|~ \rangle_u= r_1\Bigg|\begin{matrix}
s_{21}+1\\r_1-1
\end{matrix}\Bigg\rangle_u+ (1-r_1)(1-r_2)2u_1\Bigg|\begin{matrix}
t+1\\u_1-1
\end{matrix}\Bigg\rangle_u+(1-r_1)(1-r_2)3v\Bigg|\begin{matrix}
u_{2}+1\\v-1
\end{matrix}\Bigg\rangle_u
\label{qad2}
\end{align}
\item $q\cdot b_i$
\begin{align}
&\Big(q\cdot b_1\Big) |~\rangle_u= \bigg\{(1\cdot 2)(q\cdot b_1)-\tilde{N}_1 \Big(q\cdot a^\dagger_1\Big)\Big(b_1\cdot b^\dagger_2\Big)\bigg\} |s_{12}-1\rangle_u \nonumber\\
&= (1\cdot 2)^{s_{12}} {\Big(q\cdot b_1\Big)|s_{12}=0\rangle_u}+ \sum\limits_{r=0}^{s_{12}-1} (1\cdot 2)^r \bigg\{-\tilde{N}_1\Big(q\cdot a^\dagger_1\Big)\Big(b^\dagger_2\cdot b_1\Big)\bigg\} |s_{12}-1-r\rangle_u
\end{align}
Now, 
\begin{align}
&\Big(q\cdot b_1)|s_{12}=0\rangle_u=\bigg\{(2\cdot 1)\Big(q\cdot b_1\Big)+\Big(q\cdot a^\dagger_2-\tilde{N}_1\Big(q\cdot a^\dagger_1\Big)\Big(a^\dagger_2\cdot a_1\Big)\bigg\}\Big|\begin{matrix}
s_{12}=0\\s_{21}-1
\end{matrix}\Big\rangle_u\nonumber\\
&=(2\cdot 1)^{s_{21}}{\Big(q\cdot b_1\Big)|s_{ij}=0\rangle_u}+\sum\limits_{r=0}^{s_{21}-1}(2\cdot 1)^r\bigg\{\Big(q\cdot a^\dagger_2\Big)-\tilde{N}_1 \Big(q\cdot a^\dagger_1\Big)\Big(a^\dagger_2\cdot a_1\Big)\bigg\}\Big|\begin{matrix}
s_{12}=0\\s_{21}-1-r
\end{matrix}\Big\rangle_u
\end{align}
Therefore, 
\begin{align}
&\Big(q\cdot b_1\Big) |~\rangle_u= (1\cdot 2)^{s_{12}} (2\cdot 1)^{s_{21}}{\Big(q\cdot b_1\Big)|s_{ij}=0\rangle_u}\nonumber\\
&+(1\cdot 2)^{s_{12}}\sum\limits_{r=0}^{s_{21}-1}(2\cdot 1)^r\bigg\{\Big(q\cdot a^\dagger_2\Big)-\tilde{N}_1 \Big(q\cdot a^\dagger_1\Big)\Big(a^\dagger_2\cdot a_1\Big)\bigg\}\Big|\begin{matrix}
s_{12}=0\\s_{21}-1-r
\end{matrix}\Big\rangle_u\nonumber\\
&+ \sum\limits_{r=0}^{s_{12}-1} (1\cdot 2)^r \bigg\{-\tilde{N}_1\Big(q\cdot a^\dagger_1\Big)\Big(b^\dagger_2\cdot b_1\Big)\bigg\} |s_{12}-1-r\rangle_u
\end{align}
Using $(q\cdot b_1)|s_{ij=0}\rangle_u=3r_1|s_{ij}=0,r_1-1\rangle_u$ and relations (\ref{qad1}),(\ref{qad2}),(\ref{ad1a2}) and (\ref{bd1b2}), we get 
\begin{align}
(q\cdot b_1) |~ \rangle_u=&(s_{21}+\frac{s_{12}(1-r_2)}{s_{12}+s_{21}+2}+3)r_1|r_1-1\rangle_u+3vs_{21} \Bigg|\begin{matrix}
s_{21}-1,v-1\\u_2+1
\end{matrix}\Bigg\rangle_u\nonumber\\
&+2u_1s_{21}\Big(1+\frac{1}{s_{12}+s_{21}+2}\Big)\Bigg|\begin{matrix}s_{21}-1,u_1-1\\t+1\end{matrix}\Bigg\rangle_u\nonumber
\end{align}
\item $\epsilon q^\dagger b_2a^\dagger_2$
\begin{align}
&(\epsilon q^\dagger b_2 a^\dagger_2) |~\rangle = \Big[(1 \cdot 2)(\epsilon q^\dagger b_2 a^\dagger_2)+(\epsilon a^\dagger_1a^\dagger_2q^\dagger)\Big] |s_{12}-1\rangle \nonumber\\
=& (1\cdot 2)^{s_{12}} \Big(\epsilon q^\dagger b_2 a^\dagger_2\Big)|s_{12}=0\rangle + \sum\limits_{r=0}^{s_{12}-1} (1\cdot 2)^r \Big(\epsilon a^\dagger_1a^\dagger_2q^\dagger\Big) |s_{12}-1-r\rangle\nonumber\\
=& (1\cdot 2)^{s_{12}}(2\cdot 1)^{s_{21}} (q^\dagger\cdot b^\dagger_1)^{r_1}  (\epsilon q^\dagger b_2 a^\dagger_2)|r_1=0,s_{ij}=0\rangle \nonumber\\
&+ (1-t)(1-u_i)(1-r_i)(1-v) s_{12} |s_{12}-1,t+1\rangle \nonumber\\
=&r_2\Bigg|\begin{matrix}r_2-1\\u_2+1\end{matrix}\Bigg\rangle_u+ (1-t)(1-u_i)(1-r_i)(1-v) s_{12} \Bigg|\begin{matrix}s_{12}-1\\t+1\end{matrix}\Bigg\rangle 
\end{align}
Similarly, 
\begin{align}
(\epsilon q^\dagger b_1 a^\dagger_2) |~\rangle =& [(1\cdot 2)(\epsilon q^\dagger b_1 a^\dagger_2)-\tilde{N}_1(\epsilon a^\dagger_1 a^\dagger_2 q^\dagger)(b^\dagger_2\cdot b_1)]|s_{12}-1\rangle\nonumber\\
&= r_1\Bigg|\begin{matrix}r_1-1\\u_2+1\end{matrix}\Bigg\rangle_u
\end{align}
Above, we have used $(\epsilon a^\dagger_1a^\dagger_2q^\dagger)(b^\dagger_2\cdot b_1)|~\rangle_u=0$. 
\item  $\epsilon b^\dagger_1b^\dagger_2 q$
{\footnotesize
\begin{align}
&(\epsilon b^\dagger_1b^\dagger_2q)|~\rangle_u = (1\cdot 2)^{s_{12}}(2\cdot 1)^{s_{21}}(q^\dagger\cdot b^\dagger_1)^{r_1}(q^\dagger\cdot b^\dagger_2)^{r_2}(\epsilon b^\dagger_1b^\dagger_2q) |s_{ij}=0,r_i=0\rangle \nonumber\\
&=-t \Bigg|\begin{matrix}s_{12}+1,s_{21}+1\\t-1\end{matrix}\Bigg\rangle_u+2u_1\Bigg|\begin{matrix}r_1+1,s_{12}+1\\u_1-1\end{matrix}\Bigg\rangle 
+2 u_2\Bigg|\begin{matrix}r_2+1,s_{21}+1\\u_2-1\end{matrix}\Bigg\rangle+6v\Bigg|\begin{matrix}
r_1+1,r_2+1\\v-1
\end{matrix}\Bigg\rangle
\end{align}}
Above, we have used the relations $(\epsilon b^\dagger_1b^\dagger_2q)(\epsilon q^\dagger q^\dagger q^\dagger)|0\rangle=6(q^\dagger_1\cdot b^\dagger_1)(q^\dagger \cdot b^\dagger_2)|0\rangle$, $(\epsilon b^\dagger_1b^\dagger_2q)(\epsilon a^\dagger_1 q^\dagger q^\dagger)|0\rangle=2(b^\dagger_1\cdot q^\dagger)(1\cdot 2)|0\rangle$, $(\epsilon b^\dagger_1b^\dagger_2q)(\epsilon a^\dagger_1 a^\dagger_2 q^\dagger)|0\rangle=-(1\cdot 2)(2\cdot1)|0\rangle$
\item  $\epsilon q a_1b^\dagger_1$
\begin{align}
&(\epsilon q a_1 b^\dagger_1)|~\rangle_u= \bigg\{-(\epsilon  b^\dagger_1 b^\dagger_2 q)+(1\cdot 2)(\epsilon q a_1 b^\dagger_1) \bigg\}|s_{12}-1\rangle\nonumber\\
&(1\cdot 2)^{s_{12}}(\epsilon q a_1 b^\dagger_1)|s_{12}=0\rangle+\sum\limits_{r=0}^{s_{12}-1}(1\cdot 2)^r\bigg\{-(\epsilon  b^\dagger_1 b^\dagger_2 q)\bigg\}|s_{12}-1-r\rangle
\end{align}
Now,
\begin{align}
(\epsilon q a_1 b^\dagger_1)|s_{12}=0\rangle_u&=(2\cdot 1)^{s_{21}}(\epsilon q a_1 b^\dagger_1)|s_{ij}=0\rangle_u \nonumber\\
&=-4u_1\Bigg|\begin{matrix}
u_1-1\\r_1+1
\end{matrix}\Bigg\rangle_u-4t\Bigg|\begin{matrix}
t-1\\s_{21}+1
\end{matrix}\Bigg\rangle_u
\end{align}
Therefore, 
{\footnotesize
\begin{align}
(\epsilon q a_1 b^\dagger_1)|~\rangle_u&=(s_{12}+4)t\Bigg|\begin{matrix}
s_{21}+1\\t-1
\end{matrix}\Bigg\rangle_u-2(s_{12}+2)u_1 \Bigg|\begin{matrix}
r_1+1\\u_1-1
\end{matrix}\Bigg\rangle_u-2s_{12}u_2\Bigg|\begin{matrix}
s_{12}-1,s_{21}+1\\u_2-1,r_2+1
\end{matrix}\Bigg\rangle_u\nonumber\\ &-6s_{12}v\Bigg|\begin{matrix}
s_{12}-1,r_1+1\\v-1,r_2+1
\end{matrix}\Bigg\rangle_u
\end{align}}
\item $\epsilon a^\dagger_1b_2a^\dagger_2$
\begin{align}
&(\epsilon a^\dagger_1b_2a^\dagger_2) |\rangle_u = -(1+\tilde{N}_2)(\epsilon a^\dagger_1a^\dagger_2 b_2)(1 \cdot 2) |s_{12}-1\rangle_u \nonumber\\
&=(1\cdot 2)^{s_{12}}(2\cdot 1)^{s_{21}}(\epsilon a^\dagger_1a^\dagger_2 b_2) |s_{ij}=0\rangle_u = -\bigg(\frac{s_{12}+s_{21}+4}{s_{12}+s_{21}+3}\bigg)r_{2}(1-r_1)\bigg|\begin{matrix}
r_2-1\\t+1
\end{matrix}\bigg\rangle_u
\end{align}
\item $\epsilon a_1b^\dagger_1b^\dagger_2$
\begin{align}
(\epsilon a_1b^\dagger_1b^\dagger_2)|\rangle_u& = (1+\tilde{N}_1)(\epsilon b^\dagger_1b^\dagger_2a_1)|\rangle_u= (1+\tilde{N}_1)(1\cdot 2)^{s_{12}}(2\cdot 1)^{s_{21}}(\epsilon b^\dagger_1b^\dagger_2a_1)|s_{ij}=0\rangle_u \nonumber\\
&= \bigg(\frac{s_{12}+s_{21}+4}{s_{12}+s_{21}+3}\bigg)\bigg\{t(1-r_2)\bigg|\begin{matrix}
t-1,r_2+1\\s_{21}+1
\end{matrix}\bigg\rangle +u_1\bigg|\begin{matrix}
r_1+1,r_2+1\\u_1-1
\end{matrix}\bigg\rangle_u\bigg\}
\end{align}
\item $\epsilon b_1b_2a^\dagger_2$
\begin{align}
&(b_1b_2a^\dagger_2)|\rangle_u= (1+\tilde{N}_2)(\epsilon a^\dagger_2b_1b_2)|\rangle_u=(1+\tilde{N}_2)(\epsilon a^\dagger_2b_1b_2)(2\cdot 1 )|s_{21}-1\rangle_u \nonumber\\
&= (1+\tilde{N}_2)(a^\dagger_{\bar{\alpha}}\Big(\epsilon_{\alpha\beta\gamma} a^\dagger_{2\alpha}b_{1\beta}b_{2\gamma}\Big)b^\dagger_{1\bar{\alpha}}|s_{21}-1\rangle \nonumber\\
&=(1+\tilde{N}_2) \bigg\{a^\dagger_{\bar{\alpha}}\epsilon_{\alpha\beta\gamma}a^\dagger_{2\alpha}\bigg[(1-\tilde{N}^2_1)b^\dagger_{1\bar{\alpha}}b_{1\beta}+(1+\tilde{N}_1)\delta_{\bar{\alpha}\beta}-\tilde{N}_1a_{1\bar{\alpha}}a^\dagger_{1\beta}\bigg]b_{2\gamma}\bigg\}|s_{12}-1\rangle_u \nonumber\\
&=(1+\tilde{N}_2)\bigg\{(1-\tilde{N}_1^2)(2\cdot 1)(\epsilon a^\dagger_2b_1b_2)-\tilde{N}_1(a^\dagger_2\cdot a_1)(\epsilon a^\dagger_2a^\dagger_1b_2)\bigg\}|s_{21}-1\rangle_u \nonumber\\
&=(1+\tilde{N}_2)\Big[(1-\tilde{N}_1^2)(2\cdot 1)\Big]^{s_{21}}(\epsilon a^\dagger_2b_1b_2)|s_{21}=0\rangle
\label{b1b2ad21}
\end{align}
Above, we have used $(a^\dagger_2 \cdot a_1)(\epsilon a^\dagger_2a^\dagger_1b_2) |\rangle =0 $ and $b_{1\beta}b^\dagger_{1\bar{\alpha}}=(1-\tilde{N}^2_1)b^\dagger_{1\bar{\alpha}}b_{1\beta}+(1+\tilde{N}_1)\delta_{\bar{\alpha}\beta}-\tilde{N}_1a_{1\bar{\alpha}}a^\dagger_{1\beta}$ . 
Now,
\begin{align}
&(\epsilon a^\dagger_2b_1b_2) |s_{21}=0\rangle= 
 \bigg\{b^\dagger_{2\bar{\alpha}}(\epsilon_{\alpha\beta\gamma}a^\dagger_{2\alpha}b_{1\beta}b_{2\gamma})a^\dagger_{1\bar{\alpha}}+(\epsilon a^\dagger_2b_1a^\dagger_1)\bigg\} |s_{21}=0,s_{12}-1\rangle \nonumber\\
&= \bigg[(1-\tilde{N}^2_1)(1\cdot 2)(\epsilon a^\dagger_2b_1b_2)-\tilde{N}_1(b^\dagger_2\cdot b_1)(\epsilon a^\dagger_2a^\dagger_1b_2)+\epsilon a^\dagger_2 b_1a^\dagger_1\bigg]|s_{21}=0,s_{12}-1\rangle\nonumber\\
&= \Big[(1-\tilde{N}_1^2)(1\cdot 2)\Big]^{s_{12}} (\epsilon a^\dagger_2b_1b_2)|s_{ij}=0\rangle_u + \sum\limits_{r=0}^{s_{12}-1} \Big[(1-\tilde{N}_1^2)(1\cdot 2)\Big]^r (\epsilon a^\dagger_2 b_1 a^\dagger_1)\bigg|\begin{matrix}
s_{21}=0\\s_{12}-1-r
\end{matrix}\bigg\rangle_u \nonumber\\
&= \frac{2}{3}\bigg(\frac{s_{12}+3}{s_{12}+2}\bigg) \bigg[r_1r_2 \bigg|\begin{matrix}
s_{21}=0,u_2+1\\r_1-1,r_2-1
\end{matrix}\bigg\rangle-s_{12}r_1(1-r_2)\bigg|\begin{matrix}s_{21}-1,s_{12}-1\\r_1-1,t+1\end{matrix}\bigg\rangle_u\bigg]
\end{align}
Above, we have used the relation: $b_{1\beta}a^\dagger_{1\bar{\alpha}}=(1-\tilde{N}^2_1)a^\dagger_{1\bar{\alpha}}b_{1\beta}-\tilde{N}_1b_{1\bar{\alpha}}a^\dagger_{1\beta}$ in the second line. Putting this back in (\ref{b1b2ad21}), we get
\begin{align}
(b_1b_2a^\dagger_2)|\rangle_u= \bigg(\frac{s_{12}+s_{21}+3}{s_{12}+s_{21}+2}\bigg)^2 \bigg[\frac{2}{3}r_1r_2\bigg|\begin{matrix}
r_1-1,r_2-1\\u_2+1
\end{matrix}\bigg\rangle-s_{12}r_1(1-r_2)\bigg|\begin{matrix}
s_{12}-1,t+1\\r_1-1
\end{matrix}\bigg\rangle\bigg]
\end{align}
\item $\epsilon a_1b^\dagger_1a_2$
\begin{align}
&(\epsilon a_1b^\dagger_1a_2)|\rangle=-(1+\tilde{N}_1)(\epsilon b^\dagger_1a_1a_2)|\rangle = -(1+\tilde{N}_1)\bigg\{b^\dagger_{\bar{\alpha}}\Big[\epsilon_{\alpha\beta\gamma}b^\dagger_{1\alpha}a_{1\beta}a_{2\gamma}\Big](a^\dagger_2)\bigg\}|s_{21}-1\rangle \nonumber\\
&=-(1+\tilde{N}_1)\bigg\{(1-\tilde{N}_2^2)(2\cdot 1)(\epsilon b^\dagger_1 a_1a_2)-\tilde{N}_2(b^\dagger_1\cdot b_2)(\epsilon b^\dagger_1a_1b^\dagger_2)\bigg\}|s_{21}-1\rangle \nonumber\\
&=-(1+\tilde{N}_1)\bigg\{\Big[(1-\tilde{N}_2^2)(2\cdot 1)\Big]^{s_{21}}(\epsilon b^\dagger_1a_1a_2)|s_{21}=0\rangle_u +\nonumber\\
& \sum\limits_{r=0}^{s_{21}-1}\Big[(1-\tilde{N}_2^2)(2\cdot 1)\Big]^r(-\tilde{N}_2(b^\dagger\cdot b_2)(\epsilon b^\dagger_1a_1b^\dagger_2)|s_{21}-1-r\rangle\bigg\}\nonumber\\
&=-\bigg(\frac{N_1+3}{N_1+2}\bigg)\bigg(\frac{N_2+3}{N_2+2}\bigg)\bigg(\frac{N_2+2-s_{21}}{N_2+3-s_{21}}\bigg)(2\cdot 1)^{s_{21}}(\epsilon b^\dagger_1a_1a_2)|s_{21}=0\rangle_u\nonumber\\ &-\frac{(s_{12}+s_{21}+4)(s_{12}+2)s_{21}t}{(s_{12}+s_{21}+3)(s_{12}+s_{21}+2)(s_{12}+3)}\bigg|\begin{matrix}
t-1\\r_1+1
\end{matrix}\bigg\rangle_u
\label{a1bd1a21}
\end{align}
Now, 
\begin{align}
&(\epsilon b^\dagger_1a_1a_2)|s_{21}=0\rangle_u=a^\dagger_{1\bar{\alpha}}(\epsilon_{\alpha\beta\gamma}b^\dagger_{1\alpha}a_{1\beta}a_{2\gamma})b^\dagger_{2\bar{\alpha}}|s_{21}=0,s_{12}-1\rangle_u \nonumber\\
&= \bigg[(1-\tilde{N}_2^2)(1\cdot 2)(\epsilon b^\dagger_1a_1a_2)-\tilde{N}_2(a^\dagger_1\cdot a_2)(\epsilon b^\dagger_1a_1b^\dagger_2)-(\epsilon b^\dagger_1a_2b^\dagger_2)\bigg]|s_{21}=0,s_{12}-1\rangle_u\nonumber\\
&=\Big[(1-\tilde{N}_2^2)(1\cdot 2) \Big]^{s_{12}} (\epsilon b^\dagger_1a_1a_2)|s_{ij}=0\rangle_u\nonumber\\
&+\sum\limits_{r=0}^{s_{12}-1}\Big[(1-\tilde{N}_2^2)(1\cdot 2)\Big]^r\bigg\{\tilde{N}_2(a^\dagger_1\cdot a_2)(\epsilon b^\dagger_1b^\dagger_2a_1)+(1+\tilde{N}_2)(\epsilon b^\dagger_1b^\dagger_2a_2)\bigg\}\bigg|\begin{matrix}s_{21}=0\\s_{12}-1-r\end{matrix}\bigg\rangle_u \nonumber \\
&= \bigg(\frac{s_{12}+3}{s_{12}+2}\bigg)t\bigg\{\bigg(\frac{2}{3}\bigg)-\frac{s_{12}}{3(s_{12}+3)}+s_{12}\bigg\}\bigg|\begin{matrix}
s_{21}=0\\t-1,r_1+1
\end{matrix}\bigg\rangle_u+s_{12}u_2\bigg|\begin{matrix}
s_{21}=0,s_{12}-1\\u_2-1,r_1+1,r_2+1
\end{matrix}\bigg\rangle_u
\end{align}
putting this back in (\ref{a1bd1a21}), we get 
\begin{align}
(\epsilon a_1b^\dagger_1a_2)|\rangle_u&=-\bigg(\frac{s_{12}+s_{21}+4}{s_{12}+s_{21}+2}\bigg)\bigg[\frac{(s_{12}+2)s_{21}}{(s_{12}+s_{21}+3)(s_{12}+3)}+\frac{2}{3}-\frac{s_{12}}{3(s_{12}+3)}+s_{12}\bigg]t \bigg|\begin{matrix}
t-1\\r_1+1\end{matrix}\bigg\rangle_u\nonumber\\ &-\bigg(\frac{s_{12}+s_{21}+3}{s_{12}+s_{21}+2}\bigg)^2s_{12}u_2\bigg|\begin{matrix}u_2-1,s_{12}-1\\r_1+1,r_2+1
\end{matrix}\bigg\rangle_u
\end{align}
\end{enumerate}
\subsection{Norm}
Norm can be calculated recursively as follows: 
\begin{align}
&S(s,r,u,v,t)=
{}_u\langle~ |~\rangle_u={}_u\langle s_{12}-1|(a_1\cdot b_2)|s_{12}\rangle_u={\bar f}_1^{12}(s_{12}){}_u\langle s_{12}-1| s_{12}-1\rangle_u\nonumber\\
&=\bar{f}_1^{12}(s_{12}) S(s_{12}-1) \nonumber\\
&= \Big[\bar{f}_1^{12}(s_{12})\bar{f}_1^{12}(s_{12}-1) \cdots \bar{f}_1^{12}(s_{12}=1)\Big] \nonumber\\
&~~~\Big[\bar{f}_1^{21}(s_{12}=0,s_{21})\bar{f}_1^{21}(s_{12}=0,s_{21}-1)\cdots\bar{f}_1^{21}(s_{12}=0,s_{21}=1)\Big] S(s_{ij}=0,r,u,v)
\end{align}
Now, 
\begin{align}
&S(s_{ij}=0,r,u,v,t)= {}_u\langle s_{ij}=0,r,u,v,t|s_{ij}=0,r,u,v,t \rangle_u\equiv S_0\nonumber\\
&= 3(r_1+r_2)+(1-r_1)(1-r_2)[12(u_1+u_2)+36v+6t+(1-u_1)(1-u_2)(1-v)(1-t)]
\end{align}
The above factor is $6$ if $t=1$, $12$ if $u_i=1$ , $36$ if $v=1$,and $3(r_1+r_2)$ when $r_1$ or/and $r_2$ is 1 and $1$ when $r_i=u_i=v=t=0$. 
The norm is : 
	\begin{align}
S(s,r,u,v,t)=&\Big[\bar{f}_1^{12}(s_{12})\bar{f}_1^{12}(s_{12}-1) \cdots \bar{f}_1^{12}(s_{12}=1)\Big] \nonumber\\
&\Big[\bar{f}_1^{21}(s_{12}=0,s_{21})\bar{f}_1^{21}(s_{12}=0,s_{21}-1)\cdots\bar{f}_1^{21}(s_{12}=0,s_{21}=1)\Big]S_0
\label{norm}
\end{align}

\end{document}